\documentclass[11pt,twoside]{article}
\begin{document}

\begin{center}
{\Large \bf  Quantum logic networks for probabilistic and controlled teleportation  of  unknown
quantum states}\\

\vspace {0.5cm} {\large Ting Gao $^{a,b}$}\\
{\small {\it $^a$ Department of Mathematics, Capital Normal University, Beijing 100037, China\\
$^b$ College of Mathematics and Information Science, Hebei Normal University, Shijiazhuang 050016, China\\} }
\end{center}

{\small {\bf Abstract}

We present  simplification  schemes for probabilistic and controlled   teleportation of   the  unknown quantum
states of both one-particle and two-particle and construct  efficient quantum logic networks for implementing
the new schemes by means of the primitive operations consisting of single-qubit gates, two-qubit controlled-not
gates, Von Neumann measurement and classically controlled operations.  In these schemes the teleportation  are
not always successful but with certain
probability.} \\

{\parindent=0cm \bf 1. Introduction}

 Quantum teleportation [1] is commonly understood as one of the most
important aspects of quantum information. It may have applications  in quantum computer [2,3], quantum
cryptography [4-7] and quantum dense coding [8].

In 1993, Bennett et al. [1] presented a quantum method of teleportation, by which an unknown quantum pure state
of a qubit can be transmitted from one place to another by classical communication, provided that the sender
(say Alice) and the receiver (say Bob) have previously shared halves of two-qubit entangled state.   The new
method of teleportation has interested a lot of research groups. Research work on quantum teleportation was soon
widely started up, and has got great development, theoretical and experimental as well. At  present,
teleportation has been generalized to many cases [9-16], and demonstrated with the polarization photon [17] and
a single coherent mode of field [18] in the experiments [19,20]. In usual teleportation protocol, the
entanglement of two-parties is a key ingredient. In 1998, Cohen [21] and DiVincenzo et al. [22] independently
introduced the entanglement of assistance. They considered three-party states in which one of the parties, the
assistant, performs a measurement  such
 that the remaining two parties are left with on average as much entanglement as possible. Furthermore
Laustsen et al. investigated four-party states where two ~parties ~play ~the role of assistants [23].  ~As an
obvious extension of the entanglement ~of   assistance, recently Zhou et al proposed a scheme for controlled
teleporting an unknown quantum state of one-particle [24]. In their scheme a GHZ state shared by Alice, Bob and
Charlie will service as the quantum channel, and without the cooperation of Charlie the teleportation between
Alice and Bob can not be realized, i.e. the teleportation is  supervised by Charlie. This follows exactly the
same scenario as the case of the entanglement of assistance, Charlie performing a measurement to give Alice and
Bob an entangled state. The only difference is that in the scheme of Ref. [24] the two parties use that
entanglement of assistance to perform teleportation.  The importance of this scheme lies on that it can be
utilized to construct the controlled quantum channels, which may be useful to the future quantum computers.
However if the quantum channel is not the exact GHZ state form or its equivalent forms, the faithful
teleportation is impossible. In this case  unknown quantum state  can only be teleported with certain
probability [14-16].   Unfortunately in [16] many unitary transformations must be implemented. Obviously, it is
not favorable to  the
 experimental realization of teleportation.

 It is shown that all unitary operations on arbitrarily
 many bits can be decomposed into the combinations of a set of one-bit quantum gates and two-bit Controlled-Not (CNOT)
 gates [25, 26].  In terms of only single-qubit gates,
two-qubit CNOT gates, Von Neumann measurement and classically controlled operations, Gao et al. constructed an
efficient quantum logic networks for probabilistic teleportation of an unknown two-particle  state,
three-particle state  and  four-particle state  in a general form [27, 28]. Clearly the quantum logic networks
for probabilistically teleporting an unknown state will be important in realizing the teleportation scheme in
the experiment.

In this paper we will first present  simplification  schemes for probabilistic and controlled   teleportation of
  unknown quantum states of both one-particle and two-particle, and  then construct  efficient quantum logic
 networks for implementing the new scheme by means of the primitive operations consisting of single-qubit gates,
two-qubit controlled-not gates, Von Neumann measurement and classically controlled operations.

 {\parindent=0cm \bf 2. Probabilistic and controlled
 teleportation of an unknown quantum state of one-particle}

In this section we propose a  simplification  scheme for probabilistic and controlled
 teleportation of an unknown quantum state of one-particle and  also exhibit its efficient quantum logic network.

As mentioned in  Ref. [16], three people  sender "Alice", receiver "Bob" and supervisor  "Charlie" spatially
separated will be involved in the scheme of the controlled teleportation. The purpose of this scheme is to
teleport the unknown  quantum state of the particle in  Alice's place to the particle in Bob's place supervised
by Charlie. The unknown quantum state of one-particle named "particle 1" in  Alice's place which will be
teleported is
\begin{equation}
|\psi\rangle=\alpha_0|0\rangle_1 + \alpha_1|1\rangle_1
\end{equation}
with $|\alpha_0|^2 + |\alpha_1|^2=1$. Suppose  the quantum channel, which is different from the GHZ state
slightly,
\begin{equation}
\beta_0|000\rangle_{234} + \beta_1|111\rangle_{234}
\end{equation}
shared by  particles 2, 3 and 4 belonging  to Alice, Bob and Charlie respectively has been set up.  Here
$\beta_0$ and $\beta_1$ are nonzero real numbers such that $|\beta_0|^2 + |\beta_1|^2=1$ and
$|\beta_0|<|\beta_1|$.  The initial quantum state of the whole system reads
\begin{equation}
|\Phi\rangle=(\alpha_0|0\rangle_1 + \alpha_1|1\rangle_1)\otimes(\beta_0|000\rangle_{234} +
\beta_1|111\rangle_{234}).
\end{equation}

Now Alice makes a Bell state measurement on  particles 1 and 2 and  informs Bob and Charlie of the measurement
  outcome  by the classical channels. The post-measurement states  of particles 3 and 4 are collapsed into one of the
following four unnormalized states:
\begin{eqnarray}
|\phi_0\rangle_{34}&=&\alpha_0\beta_0|00\rangle_{34} + \alpha_1\beta_1|11\rangle_{34},\nonumber\\
|\phi_1\rangle_{34}&=&\alpha_0\beta_0|00\rangle_{34} - \alpha_1\beta_1|11\rangle_{34},\nonumber\\
|\phi_2\rangle_{34}&=&\alpha_0\beta_1|11\rangle_{34} + \alpha_1\beta_0|00\rangle_{34},\\
|\phi_3\rangle_{34}&=&\alpha_0\beta_1|11\rangle_{34} - \alpha_1\beta_0|00\rangle_{34},\nonumber
\end{eqnarray}
 Bob and Charlie will determine the action according to the
information received. If Charlie would like to help Bob with  the teleportation, he should send her qubit 4
through a Hadamard gate, perform a measurement on his particle using the base $\{|0\rangle_4, |1\rangle_4\}$,
 and transfer the information to Bob via a classical channel. After Charlie's operation, particles 3 and 4
  will end up in  eight possible unnormalized states:
  \begin{eqnarray}
|\varphi_0\rangle_{34}&=&(\alpha_0\beta_0|0\rangle_3+\alpha_1\beta_1|1\rangle_3)\otimes
|0\rangle_4,\nonumber\\
|\varphi_1\rangle_{34}&=&(\alpha_0\beta_0|0\rangle_3-\alpha_1\beta_1|1\rangle_3)\otimes |1\rangle_4,\nonumber\\
|\varphi_2\rangle_{34}&=&(\alpha_0\beta_0|0\rangle_3-\alpha_1\beta_1|1\rangle_3)\otimes
|0\rangle_4,\nonumber\\
|\varphi_3\rangle_{34}&=&(\alpha_0\beta_0|0\rangle_3+\alpha_1\beta_1|1\rangle_3)\otimes |1\rangle_4,\nonumber\\
|\varphi_4\rangle_{34}&=&(\alpha_1\beta_0|0\rangle_3+\alpha_0\beta_1|1\rangle_3)\otimes
|0\rangle_4,\\
|\varphi_5\rangle_{34}&=&(\alpha_1\beta_0|0\rangle_3-\alpha_0\beta_1|1\rangle_3)\otimes |1\rangle_4,\nonumber\\
|\varphi_6\rangle_{34}&=&(-\alpha_1\beta_0|0\rangle_3+\alpha_0\beta_1|1\rangle_3)\otimes
|0\rangle_4,\nonumber\\
|\varphi_7\rangle_{34}&=&(-\alpha_1\beta_0|0\rangle_3-\alpha_0\beta_1|1\rangle_3)\otimes |1\rangle_4.\nonumber
\end{eqnarray}

As soon as Bob received  Charlie's result, Bob should introduce an auxiliary particle 5 in the state
$|0\rangle_5$ and perform a unitary transformation
\begin{equation}
U_1=\left (\begin{array}{cccc}
1&0&0&0\\
0&\beta_0/\beta_1&\sqrt {1-\beta_0^2/\beta_1^2}&0\\
0&-\sqrt {1-\beta_0^2/\beta_1^2}&\beta_0/\beta_1&0\\
0&0&0&1\\
\end{array}\right )
\end{equation}
on  particles 3 and 5, under  the base $\{|00\rangle_{35}, |01\rangle_{35}, |10\rangle_{35}, |11\rangle_{35}\}$.
After that,  a
 measurement on the auxiliary qubit 5 is performed by Bob. If the result is $|1\rangle_5$, the teleportation fails.
 However, if the measurement outcome
 $|0\rangle_5$ is obtained,  the state of the particles 3 and 4  will be one of the following states
\begin{eqnarray}
&&(\alpha_0|0\rangle_3 + \alpha_1|1\rangle_3)|0\rangle_4,\nonumber\\
&&(\alpha_0|0\rangle_3 - \alpha_1|1\rangle_3)|1\rangle_4,\nonumber\\
&&(\alpha_0|0\rangle_3 - \alpha_1|1\rangle_3)|0\rangle_4,\nonumber\\
&&(\alpha_0|0\rangle_3 + \alpha_1|1\rangle_3)|1\rangle_4,\nonumber\\
&&(\alpha_1|0\rangle_3 + \alpha_0|1\rangle_3)|0\rangle_4,\\
&&(\alpha_1|0\rangle_3 - \alpha_0|1\rangle_3)|1\rangle_4,\nonumber\\
&&(-\alpha_1|0\rangle_3 + \alpha_0|1\rangle_3)|0\rangle_4,\nonumber\\
&&(-\alpha_1|0\rangle_3 - \alpha_0|1\rangle_3)|1\rangle_4,\nonumber
\end{eqnarray}
 and then  Bob  can 'fix up' his state, recovering the initial state $|\psi\rangle$ by
 applying appropriate quantum gates. For example, Bob can transform $-\alpha_1|0\rangle_3
 + \alpha_0|1\rangle_3$ into $|\psi\rangle$ by using first an $X$ and then a $Z$
gate on the particle 3, where $X=\left(%
\begin{array}{cc}
  0 & 1 \\
  1 & 0 \\
\end{array}%
\right)$ and $Z=\left(%
\begin{array}{cc}
  1 & 0 \\
  0 & -1 \\
\end{array}%
\right)$.

In the new scheme, only a collective unitary operation $U_1$ is required to achieve the probabilistic and
controlled telepotation of an unknown general quantum state $|\psi\rangle$, which is simpler than the scheme in
[16] where eight different unitary operations are needed for the same task.

Assume $\cos\frac {\theta}{2}=\frac {\beta_0}{\beta_1}$,
  then $\left (\begin{array}{cc}
\frac {\beta_0}{\beta_1}&-\sqrt {1-\frac {\beta_0^2}{\beta_1^2}}\\
\sqrt {1-\frac {\beta_0^2}{\beta_1^2}}&\frac {\beta_0}{\beta_1}\\
\end{array}\right )=R_y(\theta)=\left (
\begin{array} {cc}
\cos\frac {\theta}{2}&-\sin\frac {\theta}{2}\\
\sin\frac {\theta}{2}&\cos\frac {\theta}{2}\\
\end{array}\right )$, $A=R_y(\frac{\theta}{2})$ and $B=R_y(-\frac{\theta}{2})$.
 Unitary transformation $U_1$ can be rewritten as
\begin{equation}
U_1=C_{53}C_{35}(I\otimes B)C_{35}(I\otimes A)C_{53},
\end{equation}
where $C_{ij}$ is controlled-NOT (CNOT) gate of control qubit $i$ and target qubit $j$. The quantum circuit for
implementing this new scheme is presented in Fig.1.

Here $R=\left (\begin{array}{cc}
\beta_0&-\beta_1\\
\beta_1&\beta_0\\
\end{array}\right )$, $H=\frac{1}{\sqrt{2}}\left (\begin{array}{cc}
1&1\\
1&-1\\
\end{array}\right )$. By the first dash line  the state of particles 2, 3 and 4 is the quantum channel.
If the measurement result on the auxiliary
 qubit at the output state is $|0\rangle_a$, the teleportation is
 successful with the final state of the particle 3 being
 reconstructed as the initial state to be teleported.

 We notice that a controlled unitary operation acting on any number of
qubits followed by the measurement of the control qubit can be replaced by the measurement of the control qubit
preceding  the controlled operation. Therefore, Fig.1 can be re-expressed as Fig.2.

In Fig.2, the controlled operation can be realized locally by Bob depending on the measurement results performed
by Alice and Charlie  on their own qubits. If and only if the outcome of Alice's or Charlie's measurement is
 1, Bob can execute the Controlled-$X$ or controlled-$Z$ ($Z=HXH$). If the measurement result of the auxiliary
particle is $|1\rangle_a$, the teleportation fails. If the result is $|0\rangle_a$, the final state of particles
3 at Bob's side will be collapsed into the original state $|\psi\rangle$, which is the desired state. That is to
say, if the measurement outcome on the state of the auxiliary qubit is $|0\rangle_a$, perfect teleportation is
accomplished.

As pointed out in Ref.[16],  the probability of successful teleportation in this scheme is $2|\beta_0|^2$.

 Evidently, if  $|\beta_0|=|\beta_1|$, then  auxiliary particle  is not needed.
 In this situation  the probabilistic and controlled teleportation becomes the
 usual controlled
 one.

{\noindent\bf 3. Probabilistic and controlled teleportation of two-particle state in a general form }

A simplier protocol and quantum circuit for probabilistic and controlled teleportation of an unknown
two-particle state in a general form  are given in this section.

  The unknown quantum state of
particles 1 and 2, to be teleported in  Alice's place is as follows
\begin{equation}
|\phi\rangle_{12}=\alpha_0|00\rangle_{12} + \alpha_1|01\rangle_{12} + \alpha_2|10\rangle_{12} +
\alpha_3|11\rangle_{12},
\end{equation}
where $|\alpha_0|^2 + |\alpha_1|^2 + |\alpha_2|^2 + |\alpha_3|^2 = 1$. Suppose the quantum channel
\begin{equation}
|\varphi\rangle_{34567}=\beta_0|00000\rangle_{34567} + \beta_1|01100\rangle_{34567} +
\beta_2|10011\rangle_{34567} + \beta_3|11111\rangle_{34567},
\end{equation}
shared by  sender Alice, receiver Bob and supervisor Charlie, who are in the different places, has been set up.
Here particles 3 and 4, particles 5 and 6, and particle 7 being  controlling particle are belonging to Alice,
Bob, and Charlie, respectively. We choose $\beta_0,\beta_1,\beta_2,\beta_3$ to be  nonzero real numbers such
that
 $|\beta_0|$ is smaller  than the others and
$|\beta_0|^2+|\beta_1|^2+|\beta_2|^2+|\beta_3|^2=1$. The restriction
 on $\beta_0, \beta_1,\beta_2 $ and $\beta_3$ would guarantee the
 success of the unitary transformation stated  later.

In order to realize the teleportation, Alice should make a Bell state measurement on  particles 2 and 3 and
particles 1 and 4,  which will project
 the state of particles 5, 6 and 7 into one of  the following unnormalized  states
{\small\begin{eqnarray}
&&|\phi_0\rangle_{567}=_{14}\langle\Phi^+|_{23}\langle\Phi^+|\phi\rangle_{12}\otimes|\varphi\rangle_{34567}
=\frac{1}{2}[\alpha_0\beta_0|000\rangle_{567} + \alpha_1\beta_2|011\rangle_{567} +
          \alpha_2\beta_1|100\rangle_{567}
          + \alpha_3\beta_3|111\rangle_{567}],\nonumber\\
&&|\phi_1\rangle_{567}=_{14}\langle\Phi^-|_{23}\langle\Phi^+|\phi\rangle_{12}\otimes|\varphi\rangle_{34567}
=\frac{1}{2}[\alpha_0\beta_0|000\rangle_{567} + \alpha_1\beta_2|011\rangle_{567} -
          \alpha_2\beta_1|100\rangle_{567} - \alpha_3\beta_3|111\rangle_{567}],\nonumber\\
&&|\phi_2\rangle_{567}=_{14}\langle\Psi^+|_{23}\langle\Phi^+|\phi\rangle_{12}\otimes|\varphi\rangle_{34567}
=\frac{1}{2}[\alpha_0\beta_1|100\rangle_{567} + \alpha_1\beta_3|111\rangle_{567} +
          \alpha_2\beta_0|000\rangle_{567} + \alpha_3\beta_2|011\rangle_{567}],\nonumber\\
&&|\phi_3\rangle_{567}=_{14}\langle\Psi^-|_{23}\langle\Phi^+|\phi\rangle_{12}\otimes|\varphi\rangle_{34567}
=\frac{1}{2}[\alpha_0\beta_1|100\rangle_{567} + \alpha_1\beta_3|111\rangle_{567} -
          \alpha_2\beta_0|000\rangle_{567} - \alpha_3\beta_2|011\rangle_{567}],\nonumber\\
&&|\phi_4\rangle_{567}=_{14}\langle\Phi^+|_{23}\langle\Phi^-|\phi\rangle_{12}\otimes|\varphi\rangle_{34567}
=\frac{1}{2}[\alpha_0\beta_0|000\rangle_{567} - \alpha_1\beta_2|011\rangle_{567} +
          \alpha_2\beta_1|100\rangle_{567} - \alpha_3\beta_3|111\rangle_{567}],\nonumber\\
&&|\phi_5\rangle_{567}=_{14}\langle\Phi^-|_{23}\langle\Phi^-|\phi\rangle_{12}\otimes|\varphi\rangle_{34567}
=\frac{1}{2}[\alpha_0\beta_0|000\rangle_{567} - \alpha_1\beta_2|011\rangle_{567} -
          \alpha_2\beta_1|100\rangle_{567} + \alpha_3\beta_3|111\rangle_{567}],\nonumber\\
&&|\phi_6\rangle_{567}=_{14}\langle\Psi^+|_{23}\langle\Phi^-|\phi\rangle_{12}\otimes|\varphi\rangle_{34567}
=\frac{1}{2}[\alpha_0\beta_1|100\rangle_{567} - \alpha_1\beta_3|111\rangle_{567} +
          \alpha_2\beta_0|000\rangle_{567} - \alpha_3\beta_2|011\rangle_{567}],\nonumber\\
&&|\phi_7\rangle_{567}=_{14}\langle\Psi^-|_{23}\langle\Phi^-|\phi\rangle_{12}\otimes|\varphi\rangle_{34567}
=\frac{1}{2}[\alpha_0\beta_1|100\rangle_{567} - \alpha_1\beta_3|111\rangle_{567} -
          \alpha_2\beta_0|000\rangle_{567} + \alpha_3\beta_2|011\rangle_{567}],\nonumber\\
&&|\phi_8\rangle_{567}=_{14}\langle\Phi^+|_{23}\langle\Psi^+|\phi\rangle_{12}\otimes|\varphi\rangle_{34567}
=\frac{1}{2}[\alpha_0\beta_2|011\rangle_{567} + \alpha_1\beta_0|000\rangle_{567} +
          \alpha_2\beta_3|111\rangle_{567} + \alpha_3\beta_1|100\rangle_{567}],\\
&&|\phi_9\rangle_{567}=_{14}\langle\Phi^-|_{23}\langle\Psi^+|\phi\rangle_{12}\otimes|\varphi\rangle_{34567}
=\frac{1}{2}[\alpha_0\beta_2|011\rangle_{567} + \alpha_1\beta_0|000\rangle_{567} -
          \alpha_2\beta_3|111\rangle_{567} - \alpha_3\beta_1|100\rangle_{567}],\nonumber\\
&&|\phi_{10}\rangle_{567}=_{14}\langle\Psi^+|_{23}\langle\Psi^+|\phi\rangle_{12}\otimes|\varphi\rangle_{34567}
=\frac{1}{2}[\alpha_0\beta_3|111\rangle_{567} + \alpha_1\beta_1|100\rangle_{567} +
          \alpha_2\beta_2|011\rangle_{567} + \alpha_3\beta_0|000\rangle_{567}],\nonumber\\
&&|\phi_{11}\rangle_{567}=_{14}\langle\Psi^-|_{23}\langle\Psi^+|\phi\rangle_{12}\otimes|\varphi\rangle_{34567}
=\frac{1}{2}[\alpha_0\beta_3|111\rangle_{567} + \alpha_1\beta_1|100\rangle_{567} -
          \alpha_2\beta_2|011\rangle_{567} - \alpha_3\beta_0|000\rangle_{567}],\nonumber\\
&&|\phi_{12}\rangle_{567}=_{14}\langle\Phi^+|_{23}\langle\Psi^-|\phi\rangle_{12}\otimes|\varphi\rangle_{34567}
=\frac{1}{2}[\alpha_0\beta_2|011\rangle_{567} - \alpha_1\beta_0|000\rangle_{567} +
          \alpha_2\beta_3|111\rangle_{567} - \alpha_3\beta_1|100\rangle_{567}],\nonumber\\
&&|\phi_{13}\rangle_{567}=_{14}\langle\Phi^-|_{23}\langle\Psi^-|\phi\rangle_{12}\otimes|\varphi\rangle_{34567}
=\frac{1}{2}[\alpha_0\beta_2|011\rangle_{567} - \alpha_1\beta_0|000\rangle_{567} -
          \alpha_2\beta_3|111\rangle_{567} + \alpha_3\beta_1|100\rangle_{567}],\nonumber\\
&&|\phi_{14}\rangle_{567}=_{14}\langle\Psi^+|_{23}\langle\Psi^-|\phi\rangle_{12}\otimes|\varphi\rangle_{34567}
=\frac{1}{2}[\alpha_0\beta_3|111\rangle_{567} - \alpha_1\beta_1|100\rangle_{567} +
          \alpha_2\beta_2|011\rangle_{567} - \alpha_3\beta_0|000\rangle_{567}],\nonumber\\
&&|\phi_{15}\rangle_{567}=_{14}\langle\Psi^-|_{23}\langle\Psi^-|\phi\rangle_{12}\otimes|\varphi\rangle_{34567}
=\frac{1}{2}[\alpha_0\beta_3|111\rangle_{567} - \alpha_1\beta_1|100\rangle_{567} -
          \alpha_2\beta_2|011\rangle_{567} +\alpha_3\beta_0|000\rangle_{567}].\nonumber
\end{eqnarray}}
Here
\begin{equation}
|\Phi^\pm\rangle_{ij}=\frac {1}{\sqrt 2}(|00\rangle_{ij}\pm |11\rangle_{ij}),~~~~~~
|\Psi^\pm\rangle_{ij}=\frac{1}{\sqrt 2}(|01\rangle_{ij}\pm |10\rangle_{ij})
\end{equation}
are four Bell states of  particles $i$ and $j$ ($i=1, 2$, $j=3, 4$).

After the  Bell state measurements, Alice informs Bob and Charlie  of the
 measurement outcomes through a classical communication. For  assisting
Alice and Bob, Charlie should make a measurement
 on particle 7 using the base
$\{\frac {1}{\sqrt 2}(|0\rangle_7 + |1\rangle_7), \frac {1}{\sqrt 2}(|0\rangle_7 - |1\rangle_7)\}$  and transmit
her results to Bob over a classical communication channel, then  Bob understands exactly in which one of the
 thirty-two states in following equations
  is located particles 5 and 6.
\begin{eqnarray}
&&|\varphi_0\rangle_{56}=\alpha_0\beta_0|00\rangle_{56} + \alpha_1\beta_2|01\rangle_{56} +
                        \alpha_2\beta_1|10\rangle_{56}+ \alpha_3\beta_3|11\rangle_{56},\nonumber\\
&&|\varphi_1\rangle_{56}=\alpha_0\beta_0|00\rangle_{56} - \alpha_1\beta_2|01\rangle_{56} +
                        \alpha_2\beta_1|10\rangle_{56}- \alpha_3\beta_3|11\rangle_{56},\nonumber\\
&&|\varphi_2\rangle_{56}=\alpha_0\beta_0|00\rangle_{56} + \alpha_1\beta_2|01\rangle_{56} -
          \alpha_2\beta_1|10\rangle_{56} - \alpha_3\beta_3|11\rangle_{56},\nonumber\\
&&|\varphi_3\rangle_{56}=\alpha_0\beta_0|00\rangle_{56} - \alpha_1\beta_2|01\rangle_{56} -
          \alpha_2\beta_1|10\rangle_{56} + \alpha_3\beta_3|11\rangle_{56},\nonumber\\
&&|\varphi_4\rangle_{56}=\alpha_0\beta_1|10\rangle_{56} + \alpha_1\beta_3|11\rangle_{56} +
          \alpha_2\beta_0|00\rangle_{56} + \alpha_3\beta_2|01\rangle_{56},\nonumber\\
&&|\varphi_5\rangle_{56}=\alpha_0\beta_1|10\rangle_{56} - \alpha_1\beta_3|11\rangle_{56} +
          \alpha_2\beta_0|00\rangle_{56} - \alpha_3\beta_2|01\rangle_{56},\nonumber\\
&&|\varphi_6\rangle_{56}=\alpha_0\beta_1|10\rangle_{56} + \alpha_1\beta_3|11\rangle_{56} -
          \alpha_2\beta_0|00\rangle_{56} - \alpha_3\beta_2|01\rangle_{56},\nonumber\\
&&|\varphi_7\rangle_{56}=\alpha_0\beta_1|10\rangle_{56} - \alpha_1\beta_3|11\rangle_{56} -
          \alpha_2\beta_0|00\rangle_{56} + \alpha_3\beta_2|01\rangle_{56},\nonumber\\
&&|\varphi_8\rangle_{56}=\alpha_0\beta_0|00\rangle_{56} - \alpha_1\beta_2|01\rangle_{56} +
             \alpha_2\beta_1|10\rangle_{56} - \alpha_3\beta_3|11\rangle_{56},\nonumber\\
&&|\varphi_9\rangle_{56}=\alpha_0\beta_0|00\rangle_{56} + \alpha_1\beta_2|01\rangle_{56} +
             \alpha_2\beta_1|10\rangle_{56} + \alpha_3\beta_3|11\rangle_{56},\nonumber\\
&&|\varphi_{10}\rangle_{56}=\alpha_0\beta_0|00\rangle_{56} - \alpha_1\beta_2|01\rangle_{56} -
          \alpha_2\beta_1|10\rangle_{56} + \alpha_3\beta_3|11\rangle_{56},\nonumber\\
&&|\varphi_{11}\rangle_{56}=\alpha_0\beta_0|00\rangle_{56} + \alpha_1\beta_2|01\rangle_{56} -
          \alpha_2\beta_1|10\rangle_{56} - \alpha_3\beta_3|11\rangle_{56},\nonumber\\
&&|\varphi_{12}\rangle_{56}=\alpha_0\beta_1|10\rangle_{56} - \alpha_1\beta_3|11\rangle_{56} +
          \alpha_2\beta_0|00\rangle_{56} - \alpha_3\beta_2|01\rangle_{56},\nonumber\\
&&|\varphi_{13}\rangle_{56}=\alpha_0\beta_1|10\rangle_{56} + \alpha_1\beta_3|11\rangle_{56} +
          \alpha_2\beta_0|00\rangle_{56} + \alpha_3\beta_2|01\rangle_{56},\nonumber\\
&&|\varphi_{14}\rangle_{56}=\alpha_0\beta_1|10\rangle_{56} - \alpha_1\beta_3|11\rangle_{56} -
          \alpha_2\beta_0|00\rangle_{56} + \alpha_3\beta_2|01\rangle_{56},\nonumber\\
&&|\varphi_{15}\rangle_{56}=\alpha_0\beta_1|10\rangle_{56} + \alpha_1\beta_3|11\rangle_{56} -
          \alpha_2\beta_0|00\rangle_{56} - \alpha_3\beta_2|01\rangle_{56},\nonumber\\
&&|\varphi_{16}\rangle_{56}=\alpha_0\beta_2|01\rangle_{56} + \alpha_1\beta_0|00\rangle_{56} +
              \alpha_2\beta_3|11\rangle_{56} + \alpha_3\beta_1|10\rangle_{56},\\
&&|\varphi_{17}\rangle_{56}=-\alpha_0\beta_2|01\rangle_{56} + \alpha_1\beta_0|00\rangle_{56} -
              \alpha_2\beta_3|11\rangle_{56} + \alpha_3\beta_1|10\rangle_{56},\nonumber\\
&&|\varphi_{18}\rangle_{56}=\alpha_0\beta_2|01\rangle_{56} + \alpha_1\beta_0|00\rangle_{56} -
          \alpha_2\beta_3|11\rangle_{56} - \alpha_3\beta_1|10\rangle_{56},\nonumber\\
&&|\varphi_{19}\rangle_{56}=-\alpha_0\beta_2|01\rangle_{56} + \alpha_1\beta_0|00\rangle_{56} +
          \alpha_2\beta_3|11\rangle_{56} - \alpha_3\beta_1|10\rangle_{56},\nonumber\\
&&|\varphi_{20}\rangle_{56}=\alpha_0\beta_3|11\rangle_{56} + \alpha_1\beta_1|10\rangle_{56} +
          \alpha_2\beta_2|01\rangle_{56} + \alpha_3\beta_0|00\rangle_{56},\nonumber\\
&&|\varphi_{21}\rangle_{56}=-\alpha_0\beta_3|11\rangle_{56} + \alpha_1\beta_1|10\rangle_{56} -
          \alpha_2\beta_2|01\rangle_{56} + \alpha_3\beta_0|00\rangle_{56},\nonumber\\
&&|\varphi_{22}\rangle_{56}=\alpha_0\beta_3|11\rangle_{56} + \alpha_1\beta_1|10\rangle_{56} -
          \alpha_2\beta_2|01\rangle_{56} - \alpha_3\beta_0|00\rangle_{56},\nonumber\\
&&|\varphi_{23}\rangle_{56}=-\alpha_0\beta_3|11\rangle_{56} + \alpha_1\beta_1|10\rangle_{56} +
          \alpha_2\beta_2|01\rangle_{56} - \alpha_3\beta_0|00\rangle_{56},\nonumber\\
&&|\varphi_{24}\rangle_{56}=\alpha_0\beta_2|01\rangle_{56} - \alpha_1\beta_0|00\rangle_{56} +
          \alpha_2\beta_3|11\rangle_{56}- \alpha_3\beta_1|10\rangle_{56},\nonumber\\
&&|\varphi_{25}\rangle_{56}=-\alpha_0\beta_2|01\rangle_{56} - \alpha_1\beta_0|00\rangle_{56} -
          \alpha_2\beta_3|11\rangle_{56}- \alpha_3\beta_1|10\rangle_{56},\nonumber\\
&&|\varphi_{26}\rangle_{56}=\alpha_0\beta_2|01\rangle_{56} - \alpha_1\beta_0|00\rangle_{56} -
          \alpha_2\beta_3|11\rangle_{56} + \alpha_3\beta_1|10\rangle_{56},\nonumber\\
&&|\varphi_{27}\rangle_{56}=-\alpha_0\beta_2|01\rangle_{56} - \alpha_1\beta_0|00\rangle_{56} +
          \alpha_2\beta_3|11\rangle_{56} + \alpha_3\beta_1|10\rangle_{56},\nonumber\\
&&|\varphi_{28}\rangle_{56}=\alpha_0\beta_3|11\rangle_{56} - \alpha_1\beta_1|10\rangle_{56} +
          \alpha_2\beta_2|01\rangle_{56} - \alpha_3\beta_0|00\rangle_{56},\nonumber\\
&&|\varphi_{29}\rangle_{56}=-\alpha_0\beta_3|11\rangle_{56} - \alpha_1\beta_1|10\rangle_{56} -
          \alpha_2\beta_2|01\rangle_{56} - \alpha_3\beta_0|00\rangle_{56},\nonumber\\
&&|\varphi_{30}\rangle_{56}=\alpha_0\beta_3|11\rangle_{56} - \alpha_1\beta_1|10\rangle_{56} -
          \alpha_2\beta_2|01\rangle_{56} +\alpha_3\beta_0|00\rangle_{56},\nonumber\\
&&|\varphi_{31}\rangle_{56}=-\alpha_0\beta_3|11\rangle_{56} - \alpha_1\beta_1|10\rangle_{56}+
          \alpha_2\beta_2|01\rangle_{56} +\alpha_3\beta_0|00\rangle_{56}.\nonumber
          \end{eqnarray}

In order to make the teleportation successful, Bob needs to recover the original state $|\phi\rangle_{12}$ at
his side from the unnormalized states of particles 5 and 6. Now an auxiliary particle 8 with the initial state
 $|0\rangle_8$ is introduced by Bob.
Under  the  base  of particles 5, 6 and 8, $\{|000\rangle_{568},$ $|001\rangle_{568},$ $|010\rangle_{568},$
$|011\rangle_{568},$ $ |100\rangle_{568},$ $ |101\rangle_{568},$ $ |110\rangle_{568},$ $|111\rangle_{568}\},$
 Bob will perform  a unitary transformation
\begin{equation}
U_2=\left (
\begin{array} {cccc}
I&&&\\
&u_2&&\\
&&u_1&\\
&&&u_3\\
\end{array}\right ),
\end{equation}
where $I=\left(%
\begin{array}{cc}
  1 & 0 \\
  0 & 1 \\
\end{array}%
\right)$, $u_i=\left (
\begin{array} {cc}
\frac {\beta_0}{\beta_i}&-\sqrt {1-\frac {\beta_0^2}{\beta_i^2}}\\
\sqrt {1-\frac {\beta_0^2}{\beta_i^2}}&\frac {\beta_0}{\beta_i}\\
\end{array}\right ) (i=1, 2, 3)$, and then perform a measurement on auxiliary  particle 8. If the outcome
is $|1\rangle_8$, the teleportation fails. However,  if the outcome is $|0\rangle_8$, the teleportation is
successful and the states of the particles 5 and 6 are collapsed into
 one of the following states
\begin{eqnarray}
&&|\psi_0\rangle_{56}=\alpha_0|00\rangle_{56} + \alpha_1|01\rangle_{56} +
                        \alpha_2|10\rangle_{56}+ \alpha_3|11\rangle_{56},\nonumber\\
&&|\psi_1\rangle_{56}=\alpha_0|00\rangle_{56} - \alpha_1|01\rangle_{56} +
                        \alpha_2|10\rangle_{56}- \alpha_3|11\rangle_{56},\nonumber\\
&&|\psi_2\rangle_{56}=\alpha_0|00\rangle_{56} + \alpha_1|01\rangle_{56} -
          \alpha_2|10\rangle_{56} - \alpha_3|11\rangle_{56},\nonumber\\
&&|\psi_3\rangle_{56}=\alpha_0|00\rangle_{56} - \alpha_1|01\rangle_{56} -
          \alpha_2|10\rangle_{56} + \alpha_3|11\rangle_{56},\nonumber\\
&&|\psi_4\rangle_{56}=\alpha_0|10\rangle_{56} + \alpha_1|11\rangle_{56} +
          \alpha_2|00\rangle_{56} + \alpha_3|01\rangle_{56},\nonumber\\
&&|\psi_5\rangle_{56}=\alpha_0|10\rangle_{56} - \alpha_1|11\rangle_{56} +
          \alpha_2|00\rangle_{56} - \alpha_3|01\rangle_{56},\nonumber\\
&&|\psi_6\rangle_{56}=\alpha_0|10\rangle_{56} + \alpha_1|11\rangle_{56} -
          \alpha_2|00\rangle_{56} - \alpha_3|01\rangle_{56},\nonumber\\
&&|\psi_7\rangle_{56}=\alpha_0|10\rangle_{56} - \alpha_1|11\rangle_{56} -
          \alpha_2|00\rangle_{56} + \alpha_3|01\rangle_{56},\nonumber\\
&&|\psi_8\rangle_{56}=\alpha_0|00\rangle_{56} - \alpha_1|01\rangle_{56} +
             \alpha_2|10\rangle_{56} - \alpha_3|11\rangle_{56},\nonumber\\
&&|\psi_9\rangle_{56}=\alpha_0|00\rangle_{56} + \alpha_1|01\rangle_{56} +
             \alpha_2|10\rangle_{56} + \alpha_3|11\rangle_{56},\nonumber\\
&&|\psi_{10}\rangle_{56}=\alpha_0|00\rangle_{56} - \alpha_1|01\rangle_{56} -
          \alpha_2|10\rangle_{56} + \alpha_3|11\rangle_{56},\nonumber\\
&&|\psi_{11}\rangle_{56}=\alpha_0|00\rangle_{56} + \alpha_1|01\rangle_{56} -
          \alpha_2|10\rangle_{56} - \alpha_3|11\rangle_{56},\nonumber\\
&&|\psi_{12}\rangle_{56}=\alpha_0|10\rangle_{56} - \alpha_1|11\rangle_{56} +
          \alpha_2|00\rangle_{56} - \alpha_3|01\rangle_{56},\nonumber\\
&&|\psi_{13}\rangle_{56}=\alpha_0|10\rangle_{56} + \alpha_1|11\rangle_{56} +
          \alpha_2|00\rangle_{56} + \alpha_3|01\rangle_{56},\nonumber\\
&&|\psi_{14}\rangle_{56}=\alpha_0|10\rangle_{56} - \alpha_1|11\rangle_{56} -
          \alpha_2|00\rangle_{56} + \alpha_3|01\rangle_{56},\nonumber\\
&&|\psi_{15}\rangle_{56}=\alpha_0|10\rangle_{56} + \alpha_1|11\rangle_{56} -
          \alpha_2|00\rangle_{56} - \alpha_3|01\rangle_{56},\nonumber\\
&&|\psi_{16}\rangle_{56}=\alpha_0|01\rangle_{56} + \alpha_1|00\rangle_{56} +
              \alpha_2|11\rangle_{56} + \alpha_3|10\rangle_{56},\\
&&|\psi_{17}\rangle_{56}=-\alpha_0|01\rangle_{56} + \alpha_1|00\rangle_{56} -
              \alpha_2|11\rangle_{56} + \alpha_3|10\rangle_{56},\nonumber\\
&&|\psi_{18}\rangle_{56}=\alpha_0|01\rangle_{56} + \alpha_1|00\rangle_{56} -
          \alpha_2|11\rangle_{56} - \alpha_3|10\rangle_{56},\nonumber\\
&&|\psi_{19}\rangle_{56}=-\alpha_0|01\rangle_{56} + \alpha_1|00\rangle_{56} +
          \alpha_2|11\rangle_{56} - \alpha_3|10\rangle_{56},\nonumber\\
&&|\psi_{20}\rangle_{56}=\alpha_0|11\rangle_{56} + \alpha_1|10\rangle_{56} +
          \alpha_2|01\rangle_{56} + \alpha_3|00\rangle_{56},\nonumber\\
&&|\psi_{21}\rangle_{56}=-\alpha_0|11\rangle_{56} + \alpha_1|10\rangle_{56} -
          \alpha_2|01\rangle_{56} + \alpha_3|00\rangle_{56},\nonumber\\
&&|\psi_{22}\rangle_{56}=\alpha_0|11\rangle_{56} + \alpha_1|10\rangle_{56} -
          \alpha_2|01\rangle_{56} - \alpha_3|00\rangle_{56},\nonumber\\
&&|\psi_{23}\rangle_{56}=-\alpha_0|11\rangle_{56} + \alpha_1|10\rangle_{56} +
          \alpha_2|01\rangle_{56} - \alpha_3|00\rangle_{56},\nonumber\\
&&|\psi_{24}\rangle_{56}=\alpha_0|01\rangle_{56} - \alpha_1|00\rangle_{56} +
          \alpha_2|11\rangle_{56}- \alpha_3|10\rangle_{56},\nonumber\\
&&|\psi_{25}\rangle_{56}=-\alpha_0|01\rangle_{56} - \alpha_1|00\rangle_{56} -
          \alpha_2|11\rangle_{56}- \alpha_3|10\rangle_{56},\nonumber\\
&&|\psi_{26}\rangle_{56}=\alpha_0|01\rangle_{56} - \alpha_1|00\rangle_{56} -
          \alpha_2|11\rangle_{56} + \alpha_3|10\rangle_{56},\nonumber\\
&&|\psi_{27}\rangle_{56}=-\alpha_0|01\rangle_{56} - \alpha_1|00\rangle_{56} +
          \alpha_2|11\rangle_{56} + \alpha_3|10\rangle_{56},\nonumber\\
&&|\psi_{28}\rangle_{56}=\alpha_0|11\rangle_{56} - \alpha_1|10\rangle_{56} +
          \alpha_2|01\rangle_{56} - \alpha_3|00\rangle_{56},\nonumber\\
&&|\psi_{29}\rangle_{56}=-\alpha_0|11\rangle_{56} - \alpha_1|10\rangle_{56} -
          \alpha_2|01\rangle_{56} - \alpha_3|00\rangle_{56},\nonumber\\
&&|\psi_{30}\rangle_{56}=\alpha_0|11\rangle_{56} - \alpha_1|10\rangle_{56} -
          \alpha_2|01\rangle_{56} +\alpha_3|00\rangle_{56},\nonumber\\
&&|\psi_{31}\rangle_{56}=-\alpha_0|11\rangle_{56} - \alpha_1|10\rangle_{56}+
          \alpha_2|01\rangle_{56} +\alpha_3|00\rangle_{56},\nonumber
\end{eqnarray}
according to the outcomes of Alice's Bell state measurements and Charlie's control operation. The teleportation
can be successfully achieved with the classical information from both Alice  and Charlie and a corresponding
unitary operation  which is easy designed on the particles 5 and 6. For example we can transform
$|\psi_{19}\rangle_{56}=-\alpha_0|01\rangle_{56} + \alpha_1|00\rangle_{56} +
          \alpha_2|11\rangle_{56} - \alpha_3|10\rangle_{56},$ into the  state
$\alpha_0|00\rangle_{56}+\alpha_1|01\rangle_{56}+\alpha_2|10\rangle_{56}+\alpha_3|11\rangle_{56}$ by using first
$Z$ on the particle 6, second $Z$ on particle 5 and $X$ on particle 6. In this new scheme we only use one
unitary transformation $U_2$, which can carry out the task with the same probability as thirty-two unitary
operations in section 2
 of Ref. [16]. This will make the teleportation easily realized.

With the results in Refs. [25, 26], we show the quantum logic networks  implementing $U_2$ using only CNOT and
single qubit gates in Fig.3.

  Here $\cos\frac {\theta_i}{2}=\frac {\beta_0}{\beta_i}$,
   $u_i=R_y(\theta_i)=\left (
\begin{array} {cc}
\cos\frac {\theta_i}{2}&-\sin\frac {\theta_i}{2}\\
\sin\frac {\theta_i}{2}&\cos\frac {\theta_i}{2}\\
\end{array}\right )$, $A_i=R_y(\frac{\theta_i}{4})$ and $B_i=R_y(-\frac{\theta_i}{4})$ for $i=1, 2, 3$.

 Based on the simplification scheme  we can construct a quantum network for probabilistic and controlled
  teleportation of an unknown
two-particle state in a general form. It is illustrated in Fig.4.

Here $T=\left(\begin{array}{cc}
  1 & 0 \\
  0 & e^{{i\pi}/4} \\
\end{array}\right)$, $T^+=\left(\begin{array}{cc}
  1 & 0 \\
  0 & e^{-{i\pi}/4} \\
\end{array}\right)$, and $S=\left(\begin{array}{cc}
  1 & 0 \\
  0 & i \\
\end{array}\right)$, and
 $R_1$, $R_2$ and $R_3$ are the single qubit rotation transformations. We carefully choose $R_1$, $R_2$ and
$R_3$ to make the state of the particles 3, 4, 5, 6 and 7 to be the quantum channel
$|\varphi\rangle_{34567}=\beta_0|00000\rangle_{34567} + \beta_1|01100\rangle_{34567} +
\beta_2|10011\rangle_{34567} + \beta_3|11111\rangle_{34567}$ by the first dash line.
 If the measurement result on the auxiliary
 qubit at the output state is $|0\rangle_7$, the teleportation is
 successful with the final state of the particle 5 and 6 being
 reconstructed as the initial state to be teleported.

  Fig.4 can be re-shown as Fig.5.

As mentioned in [16], the probability of successful teleportation in this scheme is $4|\beta_0|^2$. Clearly, if
$|\beta_0|=|\beta_2|=|\beta_3|=|\beta_4|$, then the probabilistic and controlled teleportation will become the
usual  controlled teleportation, in which an  auxiliary particle   is not needed.

 From the above analysis, we can see that Alice needs to make two
 Bell state measurements. After  Alice informs  Bob and Charlie the
 result of two Bell state measurements, which will decide Bob and
 Charlie's action, with the help of the auxiliary particle and the cooperation of
 Charlie,
 Bob  can   convert the state of
 particles 5 and 6 into an exact replica of the unknown quantum
 state of particles 1 and 2, which Alice destroyed, with certain
 probability.

 In summary,  simplication schemes for probabilistic and controlled
 teleporting the unknown quantum states of one-particle and
 two-particle is proposed. The quantum logic networks for implementing this scheme is also provided.
  We hope that these schemes will be
 realized by experiment.

 \vspace {0.5cm}

{\noindent\bf Acknowledgement }

 This work was supported by
 Hebei Natural Science Foundation.

\vspace {1cm}

{\parindent=0cm \bf References } \footnotesize
\begin{tabbing}
xxxxx\=\kill
{[1]} $~~$\> C. H. Bennett,   et al., Phys. Rev. Lett.   70 (1993)  1895.\\
{[2]} $~~$\> J. I. Cirac,  P. Zoller,  Phys. Rev. Lett.  74 (1995) 4091.\\
{[3} $~~$\> A. Barenco et al.,  Phys. Rev. Lett.  74  (1995) 4083.\\
{[4]} $~~$\> A. K. Ekert,  Phys. Rev. Lett.  67  (1991) 661.\\
{[5]} $~~$\> C. H. Bennett,  Phys. Rev. Lett.  68  (1992) 3121.\\
{[6]} $~~$\> Y. Zhang,   et al., Chin. Phys. Lett.
            15 (1998) 238.\\
{[7]} $~~$\> B. S. Shi,   G. C. Guo,  Chin. Phys. Lett.   14
(1997)  521.\\
{[8]} $~~$\> C. H. Bennett,   S. J. Wiesner,  Phys. Rev. Lett.  69 (1992) 2881.\\
{[9]} $~~$\> M. Ikram, S. Y. Zhu, M. S.  Zubairy, Phys. Rev.
A62 (2000) 022307. \\
{[10]} $~~$\> W. L. Li, C. F. Li, G. C. Guo, Phys. Rev.  A61
(2000) 034301.\\
{[11]} $~~$\> V. N. Gorbachev,   A. I. Trubilko, J. Exp. Theor. Phys.  91 (2000) 894.\\
{[12]} $~~$\> H. Lu,  G. C. Guo, Phys. Lett.   A276  (2000) 209.\\
{[13]} $~~$\> B. Zeng, X. S. Liu, Y. S. Li, G. L. Long, Commun. Theor. Phys.
 38 (2002) 537.\\
{[14]} $~~$\> B. S. Shi,  et al.,  Phys. Lett.   A268 (2000) 161.\\
{[15]} $~~$\> F. L. Yan, H. G. Tan, L. G. Yang, Commun. Theor. Phys.
37 (2002) 649.\\
{[16]} $~~$\> F. L. Yan,  D. Wang, Phys. Lett. A  316 (2003) 297.\\
{[17]} $~~$\> D. Bouwmeester,  et al.,  Nature  390 (1997) 575.\\
{[18]} $~~$\> A. Furusawa, et al.,  Science  282 (1998) 706.\\
{[19]} $~~$\>  D. Boschi,  et al., Phys. Rev. Lett.   80 (1998)  1121.\\
{[20]} $~~$\> M. A. Nielsen,   et al., Nature    396 (1998)  52.\\
{[21]} $~~$\> O. Cohen, Phys. Rev. Lett. 80 (1998) 2493.\\
{[22]} $~~$\> D. P. DiVincenzo, et al., quant-ph/9803033.\\
{[23]} $~~$\> T. Laustsen, F. Verstraete, S. J. van Enk, Quant. Info. Comp. 3 (2003) 64.\\
{[24]} $~~$\> J. D. Zhou, G. Hou, S. J. Wu, Y. D. Zhang, quant-ph/0006030.\\
{[25]} $~~$\>  A. Barenco, et al., Phys. Rev. A  52 (1995) 3457.\\
{[26]} $~~$\> M. A. Nielsen, I. L. Chuang, Quantum computation and quantum information (2000), Cambridge
University Press.\\
{[27]} $~~$\> T. Gao, Z. X. Wang, F. L. Yan, Chin. Phys. Lett. 20 (2003) 2094.\\
{[28]} $~~$\> T. Gao,  F. L. Yan, Z. X. Wang,  quant-ph/0311141.
\end{tabbing}

\end{document}